%% file: main.tex
\documentclass[conference]{IEEEtran}
\IEEEoverridecommandlockouts

\usepackage{cite}
\usepackage{amsmath,amssymb,amsfonts}
\usepackage{algorithmic}
\usepackage{graphicx}
\usepackage{comment}
\usepackage{textcomp}
\usepackage[dvipsnames]{xcolor}
\usepackage[nolist]{acronym}
\input{acronyms}

\def\BibTeX{{\rm B\kern-.05em{\sc i\kern-.025em b}\kern-.08em
    T\kern-.1667em\lower.7ex\hbox{E}\kern-.125emX}}

\input{preamble_customizations}

\begin{document}

\newcommand{\todo}[1]{\textcolor{Red}{#1}}

\title{Reversing the Clock: Layout-Aware Recovery of Design Intent from Clock Distribution Networks}

\input{authors}

\maketitle

\input{sections/00_abstract}



\input{sections/01_introduction}

\input{sections/02_background}

\input{sections/03_methodology}

\input{sections/04_case_study}

\input{sections/05_discussion}

\input{sections/acknowledgement}

\bibliographystyle{IEEEtran}
\bibliography{references}

\end{document}

%% file: acronyms.tex
\begin{acronym}
    \acro{CRE}{Clock Reverse Engineering}
    \acro{VCC}{voltage at the collector}
    \acro{3D}{three-dimensional}
    \acro{2D}{two-dimensional}
    \acro{API}{application programming interface}
    \acro{ASIC}{application-specific integrated circuit}
    \acro{BFS}{Breadth-First Search}
    \acro{CDN}{clock distribution network}
    \acro{CPU}{central processing unit}
    \acro{CT}{clock tree}
    \acro{CTA}{clock tree analysis}
    \acro{CTR}{clock tree recovery}
    \acro{NA}{netlist analysis}
    \acro{NR}{netlist recovery}
    \acro{CTS}{clock tree synthesis}
    \acro{DAG}{directed acyclic graph}
    \acro{DEF}{Design Exchange Format}
    \acro{DRC}{design rule check}
    \acro{EDA}{electronic design automation}
    \acro{EEPROM}{electrically erasable programmable read-only memory}
    \acro{EMSEC}{Embedded Security Group}
    \acro{FE}{forward engineering}
    \acro{FF}{flip-flop}
    \acro{FIB}{focused ion beam}
    \acro{FPGA}{field-programmable gate array}
    \acro{FSM}{finite-state machine}
    \acro{GDSII}{Graphic Design System II}
    \acro{HDL}{hardware description language}
    \acro{HRE}{hardware reverse engineering}
    \acro{HTTP}{Hypertext Transfer Protocol}
    \acro{IC}{integrated circuit}
    \acro{IP}{intellectual property}
    \acro{IV}{current-voltage}
    \acro{LEF}{Library Exchange Format}
    \acro{LVS}{layout versus schematic}
    \acro{MPI-SP}{Max Planck Institute for Security and Privacy}
    \acro{NDR}{non-default design rule}
    \acro{NRE}{netlist reverse engineering}
    \acro{OCV}{on-chip variation}
    \acro{PCB}{printed circuit board}
    \acro{PDK}{process design kit}
    \acro{PLL}{phase-locked loop}
    \acro{RC}{resistance-capacitance}
    \acro{RE}{reverse engineering}
    \acro{RTL}{register-transfer level}
    \acro{SEM}{scanning electron microscope}
    \acro{SMT}{satisfiability modulo theories}
    \acro{SPICE}{Simulation Program with Integrated Circuit Emphasis}
    \acro{STA}{static timing analysis}
    \acro{SoC}{system on chip}
    \acro{TLM}{tree + local meshes}
    \acro{VLSI}{very-large-scale integration}
    \acro{AOI}{AND-OR-Invert}
    \acro{I/O}{Input/Output}
    \acro{MUX}{multiplexer}
    \acro{Dual-Beam FIB-SEM}{dual-beam focused ion beam-scanning electron microscope}
    \acro{GLN}{gate-level netlist}
\end{acronym}

%% file: preamble_customizations.tex
\usepackage{xcolor}      
\usepackage{tcolorbox}
\tcbuselibrary{breakable}
\tcbuselibrary{skins}

\usepackage{xurl} 
\usepackage{orcidlink}

\usepackage{subcaption}
\usepackage{siunitx}
\usepackage{hyperref} 
\usepackage{tikz}
\usetikzlibrary{shapes.geometric, arrows.meta, positioning, fit, calc}
\usepackage{hyperref}
\usepackage[normalem]{ulem}

%% file: authors.tex
\author{
\IEEEauthorblockN{
Sascha Tommasone\IEEEauthorrefmark{1}\IEEEauthorrefmark{2} \orcidlink{0009-0003-6359-1741},
Zehra Karada\u{g}\IEEEauthorrefmark{3}\IEEEauthorrefmark{2} \orcidlink{0009-0001-6863-3184},
Christopher Pawlowicz\IEEEauthorrefmark{1} \orcidlink{0009-0008-0760-9413},
Michael Green\IEEEauthorrefmark{1},\\
Bruno Machado Trindade\IEEEauthorrefmark{1},
Eunsung SEO\IEEEauthorrefmark{1} \orcidlink{0000-0001-7946-9765},
Christof Paar\IEEEauthorrefmark{2} \orcidlink{0000-0001-8681-2277},
René Walendy\IEEEauthorrefmark{2} \orcidlink{0000-0002-5378-3833},
Steffen Becker\IEEEauthorrefmark{3}\IEEEauthorrefmark{2} \orcidlink{0000-0001-7526-5597}
}
\vspace{0.3em}
\IEEEauthorblockA{\IEEEauthorrefmark{1}TechInsights Inc., Canada, \{sascha.tommasone, cpawlowicz, mgreen, btrindade, porter.seo\}@techinsights.com}
\IEEEauthorblockA{\IEEEauthorrefmark{2}Max Planck Institute for Security and Privacy, Germany, \{firstname.lastname\}@mpi-sp.org}
\IEEEauthorblockA{\IEEEauthorrefmark{3}Ruhr University Bochum, Germany, \{firstname.lastname\}@rub.de}
}

%% file: sections/00_abstract.tex
\begin{abstract}
Hardware reverse engineering supports competitive analysis and hardware assurance by recovering information about an integrated circuit (IC) from its physical implementation.
While existing techniques primarily recover the gate-level netlist, which represents logical functionality, they often overlook physical design decisions such as placement, routing, delay insertion, and interconnect optimization.
The clock distribution network encapsulates many of these decisions; however, no prior published work has recovered this network from a fabricated IC to infer design intent.
We present a layout-aware methodology that recovers and analyzes the clock distribution network by integrating a recovered gate-level netlist with layout information extracted from scanning electron microscope imagery.
Our four-phase pipeline recovers clock-tree topology, buffering, gating and switching, interconnect delay, and crosstalk-mitigation measures. 
We demonstrate our methodology on a commercial 450\,nm IC, recovering a global H-tree backbone with local X-tree-like branching, identifying an independent clock tree and global clock gating, quantifying latency, skew, and routing lengths, and confirming the absence of dedicated crosstalk mitigation.
Together, these findings let us reason about the designer's intent.
To encourage further research and support reproducibility, we release our clock tree recovery algorithm as open source.
\end{abstract}

%% file: sections/01_introduction.tex
\section{Introduction}

Competitive analysis through \acf{HRE}---recovering a high-level description of an \ac{IC} from its physical implementation to understand how it was designed---is an established practice in the semiconductor industry \cite{torrance_state_art_2009, quadir_survey_2016}.
Beyond recovering \emph{what} a circuit does, its value lies in revealing \emph{how} the design was realized: the trade-offs among performance, power, and area that distinguish one implementation from another.
Current \ac{HRE} techniques, however, predominantly recover and analyze the \ac{GLN}~\cite{azriel_survey_2021,karadag_sok_2026}, which captures only logical functionality and omits the physical design decisions encoded in the layout---cell placement, routing, delay insertion, and interconnect optimization.

The \acf{CDN} concentrates such decisions like few other structures on a chip: it is among the most carefully engineered networks in an \ac{IC} and a dominant consumer of dynamic power, accounting for up to \qty{40}{\percent}~\cite{pangjun_low_power_2002}, so its topology, buffering, and gating closely track a design's performance and power strategy.
Yet, no prior published work has reconstructed the \ac{CDN} from a physical \ac{IC} to recover design intent.
These design decisions make a recovered \ac{CT}---unlike the netlist---a rich source for such analysis: its topology can be matched against reference designs to attribute implementations or localize third-party \acs{IP} cores, while metal-routing lengths, interconnect propagation delays, and physical countermeasures such as crosstalk shielding or buffer sizing provide further distinguishing signatures.

In this paper, we present a method to recover \acp{CDN} from reverse-engineered \acp{IC} by combining a gate-level netlist with layout information obtained from \acf{SEM} imagery, and demonstrate its effectiveness in a real-world case study.
Our contributions are as follows:

\begin{itemize}
    \item \textbf{A four-phase pipeline} that reconstructs and analyzes the \ac{CT} by combining the netlist with layout information, proceeding through netlist and layout recovery, clock-tree recovery, structural analysis, and timing analysis.
    \item \textbf{A case study} on a commercial \ac{IC} fabricated in a \qty{450}{\nano\meter} node, where layout-aware analysis recovers the \ac{CT}'s hybrid H-tree/X-tree architecture, quantifies its clock latencies, skews, and routing lengths, and detects the absence of dedicated crosstalk mitigation---enabling us to reason about the designer's intent.
    \item \textbf{An open-source artifact} comprising the algorithm for recovering the \ac{CT} graph as a plugin for the \textit{HAL} framework~\cite{embedded_security_group_hal_2019}, and a benchmark netlist; its setup and execution scripts are permanently hosted on Zenodo.\footnote{
    \hangindent=1.45em\hangafter=1
    \url{https://doi.org/10.5281/zenodo.21263979}.
    As the case-study netlist cannot be published, the artifact utilizes the open-source \textit{Cloneless1} netlist~\cite{moos_cloneless1} to remain runnable.}
\end{itemize}

%% file: sections/02_background.tex
\section{Background}

\subsection{Clock Distribution}
The physical clock distribution implementation reflects how designers balance performance, area, and power through placement, routing, delay insertion, and interconnect optimization~\cite{friedman_clock_2001}. 
Recovering this design intent requires reverse engineering the physical implementation and the circuit's functional description.

\subsubsection{Clock Distribution Networks}
Clock signals are distributed through carefully engineered structures called \acfp{CDN} or \acfp{CT}~\cite{xanthopoulos_clocking_2009}. 
Various topologies trade off latency, power, area, skew, jitter, and scalability~\cite{yeh_clock_2006}; the H-tree and X-tree are the most widely adopted~\cite{friedman_clock_2001, xanthopoulos_clocking_2009}. 
The H-tree uses symmetric, recursive H-shaped branching to keep all sinks equidistant from the source, theoretically eliminating skew, though process variations introduce minor mismatches. 
The X-tree instead uses diagonal X-shaped branches to reduce wire length and latency, but is harder to balance and more sensitive to layout-induced skew.

\subsubsection{Clock Integrity Measures}
To counter clock skew, clock buffers and inverters can balance arrival times and regenerate the signal, at the cost of area and switching power~\cite{chen_clock_2010, pangjun_low_power_2002}.
To preserve signal integrity, they protect the frequently switching clock lines against \textit{crosstalk} from capacitively coupled neighbors using wider spacing or grounded guard tracks, consuming routing resources~\cite{vittal_crosstalk_1999, razif_mitigation_2019}.
These choices manifest physically---as inserted cells, wire detours, and spacing patterns---and remain readable in the layout.

\subsection{Hardware Reverse Engineering}
To infer design intent from a physical \ac{IC}, both the \acf{GLN} for \ac{CTR} and detailed layout information for \ac{CTA} are required. 
These are obtained through \ac{NR}, the initial phase of \ac{HRE}~\cite{quadir_survey_2016, torrance_state_art_2009, karadag_sok_2026}. 
During \ac{NR}, the \ac{IC} layers are exposed and imaged, then stitched, aligned, and segmented into a coherent \ac{3D} layout representation, from which standard cells and connectivity are identified, producing the \ac{GLN}~\cite{lippmann_integrated_2019}. 
Because \ac{NR} is error-prone, it introduces opens and shorts into both artifacts, which \ac{CTR} must tolerate.
Subsequent \ac{NA} conventionally operates only on the \ac{GLN}~\cite{azriel_survey_2021, subramanyan_reverse_2014}, reducing \acp{CDN} to a graph of interconnected gates. 
However, distinguishing information about different implementations, such as topology, routing, spacing, and crosstalk control, is captured only in the recovered layout but is set aside by conventional analysis, leaving a rich source of information untapped.

%% file: sections/03_methodology.tex
\section{Methodology}
To reverse engineer \acp{CDN} and ultimately uncover design intent, we implement a four-phase pipeline (see \autoref{fig:methodology}). 
The process begins by recovering the \ac{IC} layout and \acf{GLN}, followed by recovering the \ac{CT}. 
The final two phases, structural and timing analyses, then yield the insights from which we infer design intent.

\input{figures/4_phases_flow_diagram}

\subsection{Phase 1: Netlist \& Layout Recovery}
We cross-section the \ac{IC} to guide the subsequent delayering and to support later parasitic extraction and \ac{3D} modeling.
During delayering, we image each exposed layer with a \ac{SEM} and segment the images into a coherent \ac{3D} polygonal layout, and then preprocess it (dilation, per-layer unary union, erosion, simplification) to close gaps and reduce geometric noise. 
We then recover the \ac{GLN} from this layout by reverse-engineering standard cells and signal-tracing interconnectivity.

We refine this recovery iteratively, using the netlist as the correctness signal: basic \acp{DRC} flag multi-driven nets, undriven nets, and gates driving power or ground; the violations guide targeted layout corrections (fixing shorts and opens, rerunning signal tracing, adding cells), repeated until the netlist stabilizes. 
These checks are deliberate---undriven nets mark termination points for \ac{CTR}, multi-driven nets often signal shorts that divert the clock into datapath or control logic, and power/ground-driving gates are rare enough to warrant manual inspection. 
The resulting cross-sectional images, layout, and \ac{GLN} form the basis for all subsequent phases.

\subsection{Phase 2: Clock Tree Recovery}
The recovered \ac{GLN} lacks structural and semantic information available during the original design process, such as module hierarchy and signal names~\cite{karadag_sok_2026}. 
Consequently, the \ac{CT} and its associated clock signals are typically unknown.
In the second phase, our novel \ac{CTR} algorithm reconstructs the \ac{CT} solely from recovered data, combining structural netlist properties with layout information. 
The algorithm is released as an open-source plugin\footnote{\url{https://github.com/emsec/hal/tree/master/plugins/clock_tree_extractor}} for HAL~\cite{embedded_security_group_hal_2019}.
It comprises three steps: integrating layout information with the netlist to identify global inputs and outputs, identifying clock-dividing structures, and constructing the \ac{CT}.

\subsubsection{Identifying Global Inputs}
A \ac{CT} may comprise multiple subtrees driven by a common global clock input. 
When recovery relies solely on the \ac{GLN}, these subtrees appear disconnected because their roots terminate at an unidentified global input, making them resemble undriven wires. 
As a result, the recovered \ac{CT} becomes partitioned since the \ac{CTR} algorithm cannot distinguish between externally-driven wires and \ac{NR} errors. 
Therefore, we identify global inputs before \ac{CTR} to resolve this ambiguity and assemble the subtrees into a complete \ac{CT}.

Global wires traverse the digital-core boundary.
We identify them by approximating the boundary as a polygon and selecting metal-layer polygons that intersect this boundary. 
During \ac{NR}, signal tracing organizes polygons into wires, referred to as nets within the context of \acp{GLN}. 
Consequently, boundary-crossing polygons can be directly mapped to their corresponding \ac{GLN} nets. 
We then classify each global net heuristically as a global input if it lacks a driver in the \ac{GLN}, and as a global output otherwise.

\subsubsection{Identifying Clock Dividers}
Flip-flops are generally sinks in the \ac{CDN}, with their clock input driven by the \ac{CT}~\cite{xanthopoulos_clocking_2009}.
However, when used within a clock divider, flip-flops function as active elements in the clock path and must be represented as such in the \ac{CT}.
The toggle flip-flop, which halves the clock frequency, is the most basic example: it is identifiable directly from the netlist, as its inverted output connects back to its own data input.
Before \ac{CT} construction, we detect all such toggle flip-flops so they can be treated as active clock elements rather than sinks during traversal.

\subsubsection{Recovering the Clock Tree}
We adopt a bottom-up approach where we construct the \ac{CT} by tracing from sequential gates toward the clock sources. 
This strategy does not require a priori knowledge of clock sources and is more robust to local recovery errors than top-down \ac{CT} construction.

The \ac{CT} construction process begins by initializing a \ac{DAG}, hereafter referred to as the \ac{CT} graph, with all flip-flops as nodes. 
From each flip-flop, a modified \ac{BFS} is performed on the \ac{GLN} to traverse the clock network in the reverse direction. 
At each step, the algorithm recursively follows valid fan-in connections. 
In contrast, control inputs (such as set, reset, and multiplexer select signals), power and ground nets, and invalid connections (such as unrouted or multi-driven nets) are excluded. 
Global inputs, buffers, inverters, and delay gates are treated as active clock elements and thus serve as reference points during traversal to stay on the clock path. 
A branch is terminated upon reaching a latch or a non-clock-dividing flip-flop, as these sequential elements are typically part of the control logic that steers clock gates rather than the \ac{CT}. 
If a branch cannot be traversed further and does not terminate at a reference point, it is discarded because it cannot be reliably associated with the \ac{CT}. 
If a traversal terminates at a reference point, the gates along the clock path are merged into the \ac{CT} graph. 
Subsequently, edges are inserted in the order of traversal, and nodes without predecessors are designated as roots.

Having recovered the \ac{CT} with its global inputs and outputs and clock dividers, we can now analyze its structure in depth.

\subsection{Phase 3: Clock Tree Structural Analysis}
Based on the output from phases 1 and 2---the \ac{CT} graph, the \ac{GLN}, and the physical layout---the structural analysis recovers all \ac{CT} branches. It examines the implemented topology, routing, buffering strategies, dynamic power-reduction techniques, and signal-integrity measures to infer underlying design principles.
As part of the analysis, we developed the following algorithms:

\begin{itemize}
\item An \textit{undirected graph-based layout representation} enables efficient intersection queries and short detection.
The graph nodes represent layout polygons, and edges denote intersections across the same or adjacent layers; it is built during signal tracing.
\item A \textit{ray-casting method} identifies adjacent polygons and computes spacing profiles within a layer. 
This approach samples points along each polygon edge and extends probes along the corresponding outward edge normal until they encounter an adjacent polygon or reach a predefined maximum distance.
\item A \textit{centerline-graph method} estimates track and total wire lengths by finding the shortest path between two points on the same wire within the layout graph. 
It constructs a centerline graph for each traversed metal polygon and uses adjacent via nodes as entry and exit points to estimate each segment's length. 
\end{itemize}

These algorithms provide the foundation for all subsequent layout-based analyses.

\subsubsection{Identifying the Topology}
To infer the implemented topology, we integrate the \ac{CT} graph with gate placement coordinates to produce a layout-aware visualization, since the \ac{CT} graph alone does not capture physical implementation details.
We visualize the \ac{CT} graph at different tree-traversal depths from each root to help identify multiple combined \ac{CT} topologies.

\subsubsection{Branch Recovery}
\label{sec:branch_recovery}
To perform the subsequent analyses, we must identify the various \ac{CT} branches, each representing a distinct clock signal path to a sink, from the recovered \ac{CT} graph.
We extract these branches by traversing the \ac{CT} graph from each root to every reachable sink, recording each unique root-to-sink path as an ordered sequence of triples: driver, interconnect segment, and load.

\subsubsection{Identifying Buffering}
Clock buffers and clock inverters serve as the primary active elements in a \ac{CT}~\cite{friedman_clock_2001}. 
Analyzing their placement and distribution can provide insights into active delay insertion, area optimization, and slew-rate recovery. 
The buffering strategy can be visualized using either the previous \ac{CT}-topology approach or a comprehensive \ac{3D} interconnect model. 
In topology-based visualization, we highlight clock buffers and inverters using distinct colors. 
In contrast, we construct the \ac{3D} model by extruding layout geometry using the layer heights extracted from the cross-sectional images and overlaying the active elements onto the resulting model.
Compared with the \ac{2D} topology, the \ac{3D} visualization offers the reconstructed physical clock-tree implementation for inspection and is easier to analyze.

\subsubsection{Identifying Dynamic Power Reduction}
Given that \acp{CT} can contribute up to $\qty{40}{\percent}$ of dynamic power consumption~\cite{pangjun_low_power_2002}, modern \acp{IC} frequently implement clock gating and clock switching to minimize dynamic power or facilitate operation across multiple clock domains~\cite{farrahi_activity_driven_2001, zeidler_glitch_free_2020}. 

Clock gating is commonly implemented using combinational gates in the \ac{CT} that are not buffers, inverters, or multiplexers. 
In contrast, clock switching is generally realized using multiplexers or similar logic structures. 
These distinctive configurations enable effective identification within the \ac{CT} graph. 
Functional analysis of these elements provides the conditions for gating or switching clock signals, while \ac{NA} reveals the controlling logic and the components affected.

\subsubsection{Identifying Routing}
\Ac{CT} routing complexity is determined by two primary factors: the metal layers utilized and the number of layer transitions per branch relative to the total branch wire length. 
We determine metal-layer utilization by automatically estimating the physical length of each \ac{CT} track and summing these lengths across all metal layers using our centerline-graph-based distance algorithm. 
We calculate the total branch wire length by summing the lengths of all metal-track segments that form a branch, while layer transitions are quantified by counting the via-layer crossings along the branch.
However, we exclude standard-cell interconnects from both metrics because they are not part of the \ac{CT} routing. 
Specifically, we first identify the metal layers reserved for standard-cell routing and then exclude all \ac{CT} metal polygons on these layers that intersect a standard-cell pin and overlap the corresponding standard-cell boundary by more than $\qty{50}{\percent}$.

\subsubsection{Determining Signal-Integrity Measures}
\label{sec:crosstalk_detection}
Determining whether shielding or increased-spacing crosstalk control is implemented globally, locally, or not at all requires both the \ac{CT} graph and layout information.
First, we generate a spacing profile for each metal track in the recovered \ac{CT}, recording its surrounding tracks and their respective spacings only within the same layer.
We then apply method-specific post-processing to these spacing profiles to infer the implemented crosstalk mitigation strategy and classify it as global, local, or absent.

\subsection{Phase 4: Clock Tree Timing Analysis}
\label{sec:timing_method}
Structural analysis does not capture the dynamic behavior of a clock signal. 
Understanding system performance therefore requires evaluating its timing characteristics.

Each \ac{CT} branch exhibits a unique clock latency, determined by interconnect and standard-cell propagation delays~\cite{xanthopoulos_clocking_2009}. 
Accordingly, we build a SPICE circuit model for each branch, simulate it, and use the resulting latencies to compute clock skews. 
We develop these branch models by representing each interconnect segment as a distributed \ac{RC} circuit and connecting them using the corresponding SPICE standard-cell models.
The standard cells can be approximated by measuring transistor characteristics, such as gate length and width, and utilizing generic SPICE transistor models to build a standard-cell model.
To estimate metal-track resistances from the layout and cross-sectional imagery, we adopt the widely used \textit{sheet-resistance model}~\cite{baker_cmos_2010}, and to estimate metal-track capacitances, we use the open-source \ac{3D} field solver FasterCap~\cite{fastfieldsolvers_srl_fastercap_2019}.
We construct the \ac{3D} payloads for \textit{FasterCap} by integrating the layout geometry with layer heights obtained from cross-sectional imagery.

%% file: figures/4_phases_flow_diagram.tex
\definecolor{cteal}{RGB}{29,158,117}
\definecolor{cpurple}{RGB}{127,119,221}
\definecolor{camber}{RGB}{186,117,23}
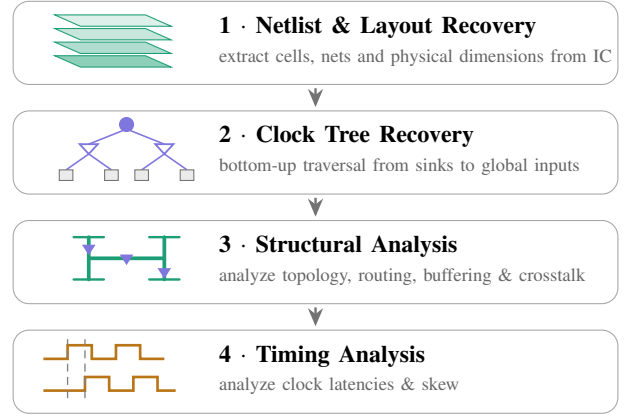
\begin{figure}[t]
\centering
\begin{tikzpicture}[
  x=1cm, y=1cm,
  panel/.style={draw=black!35, rounded corners=6pt},
  flow/.style={-{Stealth[length=2.4mm,width=2mm]}, black!55, line width=0.9pt},
  ptitle/.style={anchor=west, font=\small\bfseries},
  psub/.style={anchor=west, font=\scriptsize, black!60, align=left},
  wire/.style={cteal, line width=1.4pt, line cap=round},
  twig/.style={cteal, line width=0.9pt, line cap=round},
  gedge/.style={cpurple, line width=0.6pt},
]
\begin{scope}[shift={(0,0)}]
  \draw[panel] (0,0) rectangle (8,1.1);
  \foreach \i/\op in {0/15, 1/25, 2/35, 3/50}{
    \draw[cteal, fill=cteal!\op]
      (0.55,0.92-0.18*\i) -- ++(1.20,0) -- ++(0.34,-0.15)
      -- ++(-1.20,0) -- cycle;
  }
  \node[ptitle] at (2.6,0.76) {1 $\cdot$ Netlist \& Layout Recovery};
  \node[psub]   at (2.6,0.36) {extract cells, nets and physical dimensions from \ac{IC}};
\end{scope}
\draw[flow] (4,-0.05) -- (4,-0.30);
\begin{scope}[shift={(0,-1.45)}]
  \draw[panel] (0,0) rectangle (8,1.1);
  \fill[cpurple] (1.5,0.92) circle (1.0mm);                    
  \draw[gedge] (1.5,0.86) -- (1.0,0.66)  (1.5,0.86) -- (2.0,0.66);
  \foreach \x in {1.0, 2.0}{                                    
    \draw[cpurple, line width=0.7pt]
      (\x-0.12,0.66) -- (\x+0.12,0.66) -- (\x,0.53) -- cycle;
    \draw[gedge] (\x,0.53) -- (\x-0.30,0.32)  (\x,0.53) -- (\x+0.30,0.32);
  }
  \foreach \x in {0.70, 1.30, 1.70, 2.30}{                      
    \draw[black!55, fill=black!8] (\x-0.09,0.18) rectangle (\x+0.09,0.32);
  }
  \node[ptitle] at (2.6,0.76) {2 $\cdot$ Clock Tree Recovery};
  \node[psub]   at (2.6,0.36) {bottom-up traversal from sinks to global inputs};
\end{scope}
\draw[flow] (4,-1.50) -- (4,-1.75);
\begin{scope}[shift={(0,-2.90)}]
  \draw[panel] (0,0) rectangle (8,1.1);
  \draw[wire] (1.0,0.60) -- (2.0,0.60);                         
  \draw[wire] (1.0,0.32) -- (1.0,0.88)  (2.0,0.32) -- (2.0,0.88); 
  \foreach \x in {1.0, 2.0} \foreach \y in {0.32, 0.88}
    \draw[twig] (\x-0.20,\y) -- (\x+0.20,\y);                   
  \foreach \x/\y in {1.50/0.65, 1.00/0.78, 2.00/0.47}{          
    \fill[cpurple] (\x-0.09,\y) -- (\x+0.09,\y) -- (\x,\y-0.14) -- cycle;
  }
  \node[ptitle] at (2.6,0.76) {3 $\cdot$ Structural Analysis};
  \node[psub]   at (2.6,0.36) {analyze topology, routing, buffering \& crosstalk};
\end{scope}
\draw[flow] (4,-2.95) -- (4,-3.20);
\begin{scope}[shift={(0,-4.35)}]
  \draw[panel] (0,0) rectangle (8,1.1);
  \draw[camber, line width=1pt]                                 
    (0.40,0.72) -- (0.72,0.72) -- (0.72,0.90) -- (1.04,0.90) -- (1.04,0.72)
    -- (1.36,0.72) -- (1.36,0.90) -- (1.68,0.90) -- (1.68,0.72) -- (2.00,0.72);
  \draw[camber, line width=1pt]                                 
    (0.40,0.30) -- (0.95,0.30) -- (0.95,0.48) -- (1.27,0.48) -- (1.27,0.30)
    -- (1.59,0.30) -- (1.59,0.48) -- (1.91,0.48) -- (1.91,0.30) -- (2.15,0.30);
  \draw[black!50, densely dashed, line width=0.5pt]
    (0.72,0.20) -- (0.72,0.98)  (0.95,0.20) -- (0.95,0.98);
  \node[ptitle] at (2.6,0.76) {4 $\cdot$ Timing Analysis};
  \node[psub]   at (2.6,0.36) {analyze clock latencies \& skew};
\end{scope}
\end{tikzpicture}
\caption{Our four-phase pipeline, from a physical \ac{IC} to a reconstructed and analyzed \ac{CT} that grounds design-intent recovery\@.}
\label{fig:methodology}
\vspace{-8pt}
\end{figure}

%% file: sections/04_case_study.tex
\section{Case Study}
\label{sec:case_study}
The case study applies the methodology to the digital core of a real-world \ac{IC}, fabricated in a $\qty{450}{\nano\meter}$ technology node and operating at $\qty{2}{\mega\hertz}$ or $\qty{3}{\mega\hertz}$ depending on the supplied voltage.
Its three metal layers, M1 to M3, have approximate track widths of $\qty{625}{\nano\meter}$, $\qty{395}{\nano\meter}$, and $\qty{475}{\nano\meter}$, respectively.

\begin{figure}[htb]
\centering
\includegraphics[width=\linewidth]{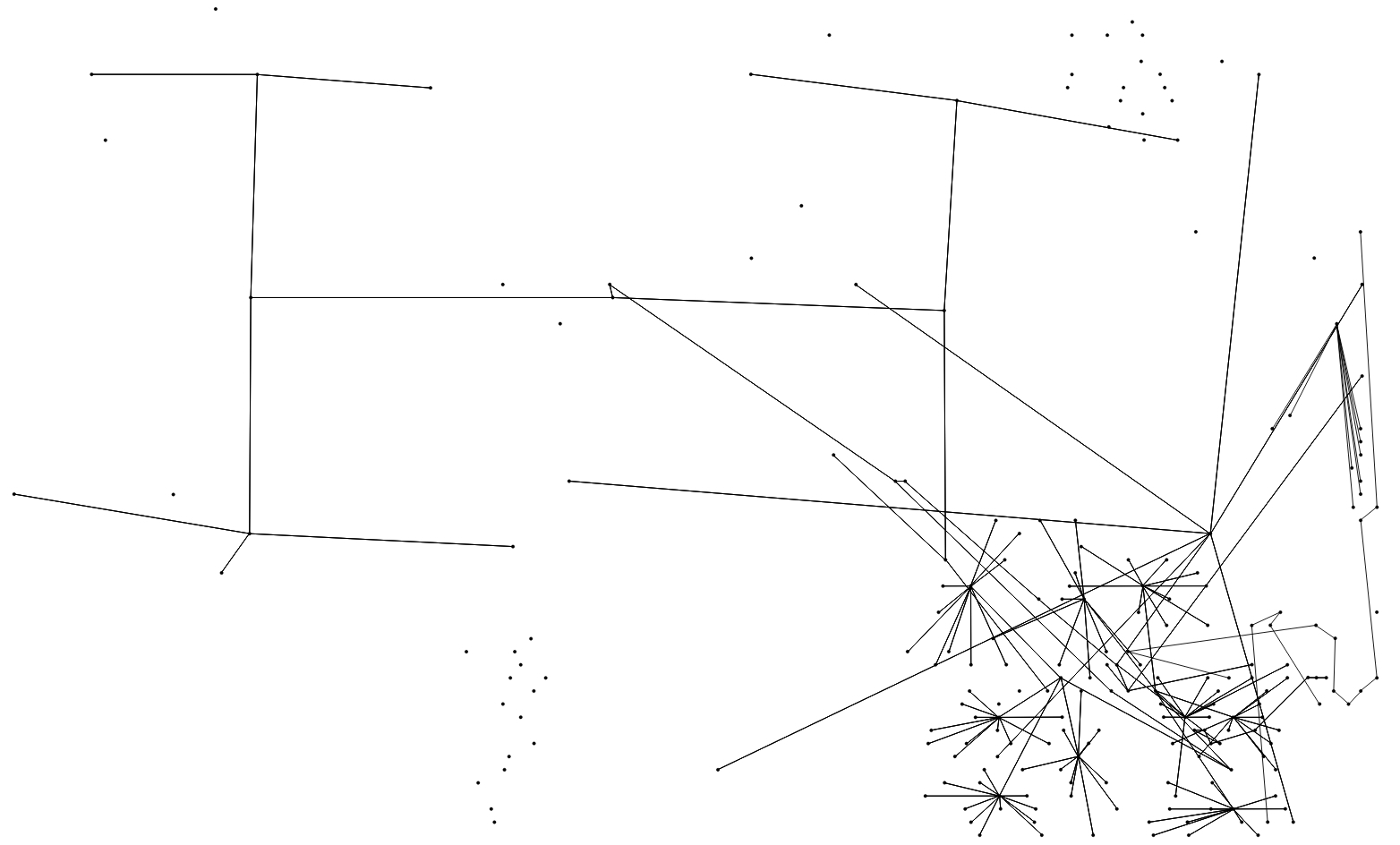}
\caption{
Layout-aware \ac{2D} visualization of an excerpt of the \ac{CT}, constrained to a maximum traversal depth of 11\@, showing an H-tree structure with the
I\textsuperscript{2}C subtree in the bottom-right\@.
}
\label{fig:clock_tree_2D}
\vspace{-8pt}
\end{figure}

The \ac{CDN} of the digital core has three global inputs: two clock sources and a select signal that determines which source is used. 
The selected clock is distributed to a partial I\textsuperscript{2}C interface consisting of X-tree-like sinks and, through an H-tree backbone, to other X-tree-like sinks.
The branches to these destinations are controlled by additional logic that determines whether the I\textsuperscript{2}C module receives its clock through the path shared with the H-tree backbone enable, during normal operation, or through a dedicated direct path that can remain active, whereas the backbone is disabled (see \autoref{fig:architecture}).
The independent I\textsuperscript{2}C lets the interface stay responsive while the rest of the core is gated. 
This architecture can reduce power consumption by disabling the main clock distribution while keeping the I\textsuperscript{2}C interface operational to monitor external events.
\input{figures/branches_overview}
The H-tree backbone forms the primary clock distribution structure. 
Its symmetric topology minimizes clock skew at the expense of increased routing resources, while naturally matching the preferred routing directions of the upper metal layers.
At lower levels, symmetry is gradually relaxed to improve routing efficiency and accommodate the irregular placement of sink clusters.
\autoref{fig:clock_tree_2D} shows a layout-aware visualization of the I\textsuperscript{2}C sinks connected to an H-tree.
Previous \ac{NA} already revealed the presence of this interface, but didn't provide information about its specific clocking.

The third subtree is a complete yet functionally deactivated divider-based clock-switching structure in the top-right region.

\subsection{The Recovered Clock Tree as Basis for Analysis}
\label{sec:foundation}
The completeness of the recovered \ac{CT} can be measured by the number of flip-flop coverages, since each flip-flop requires a clock.
The recovered \ac{CT} covers $1,290$ of the $1,346$ flip-flops ($\qty{95}{\percent}$). 
The remaining $56$ flip-flops are distributed across $50$ small splinters, mostly resulting from netlist errors such as opens or shorts.
Covering the vast majority of expected flip-flops is sufficient to reason about the design principles behind the clock distribution.
Therefore, this case study focuses on the largest component.
Within this component, three global inputs act as the root.
Two are clock sources, and one serves as a clock select.
All three converge at a clock-switching structure at the head of the \ac{CT}, where the select signal determines which of the clock sources should be used.

\subsection{Clock Control Structures}
\label{sec:control}
Most elements of the recovered \ac{CT} serve only to distribute the clock: the metal interconnect, buffers, and inverters merely propagate the clock signal.
Along these branches, however, reside four active control structures---the only points at which the clock distribution can be reconfigured during operation.
They determine which source drives the tree, whether the clock reaches its sinks at all, and at which frequency.

\paragraph{Clock Source Selection}
The physical root of the recovered \ac{CT} is a compound \acs{AOI} gate that functions as a two-input \acs{MUX}.
This \acs{MUX} is the clock-switching structure at which the three global inputs converge: the two clock sources form its data inputs, and the select signal its control input.
Depending on the select signal, the recovered \ac{CT} is driven either by the on-chip clock or by an external clock supplied through an \acs{I/O} cell.
An \acs{I/O} cell also supplies the select signal.
Every sink in the \ac{CT} therefore operates on whichever clock this selection admits.
In the analyzed configuration, this structure always selects the on-chip clock source.

\paragraph{H-Tree Backbone Enable}
Downstream of this source selection, an active-low enable clock gate, implemented as an \textbf{OR2} gate, is positioned directly upstream of the H-tree backbone.
In addition to the clock, this gate receives a control signal; when asserted high, it blocks all clock edges, preventing the entire downstream H-tree from receiving a clock.
The affected flip-flops retain their state but cease switching, thereby eliminating the clock-related dynamic power of most of the digital core.
A single gate with this reach constitutes coarse-grained clock gating: rather than gating individual registers or modules, the designers provide a single global switch.

\paragraph{I\textsuperscript{2}C Clocked Alone}
While the backbone is disabled, the I\textsuperscript{2}C \ac{CT} can remain clocked: its root is a two-input \acs{MUX} that dynamically selects the clock driving the downstream independent \ac{CT}.
One input is derived directly from the recovered \ac{CT} root and is thus unaffected by the backbone enable; the other is obtained after the H-tree backbone clock gate, before branching into the H-tree.
These two inputs enable a potential low-power mode: the gated input keeps the interface synchronized with the core under normal operation, whereas the ungated input allows the I\textsuperscript{2}C interface to monitor external events. 
Meanwhile, the remainder of the device is shut down.

\paragraph{Clock Division and Selection}
An 8-to-1 \acs{MUX} in the top-right region of the digital core selects among eight clock divisions generated by a chain of toggle flip-flops. 
This structure taps the on-chip clock before the root selector, bypassing both source selection and the backbone enable. 
The divisions span from $2^{-7}$ to $2^{-15}$ of the primary clock, except $2^{-13}$. 
In the configuration under analysis, the select inputs are tied high, selecting $2^{-15}$. 
This clock signal drives a 10-bit shift register, whose output is bypassed by a 2-to-1 \acs{MUX}, rendering the structure inactive. 
The presence of a complete yet disabled implementation indicates a provisioned capability, potentially intended for manufacturing test, debugging, or other configurable features.

\subsection{Physical Implementation}
\label{sec:silicon}
The control structures let the I\textsuperscript{2}C \ac{CT} and the H-tree backbone operate independently, yet their physical implementation must still keep both parts synchronized during normal operation.
We therefore examine logic depth, metal routing, buffering, and crosstalk control of the recovered \ac{CT}.

\subsubsection{Groups Revealed by Depth}
\begin{figure}[t]
\centering
\includegraphics[width=\linewidth]{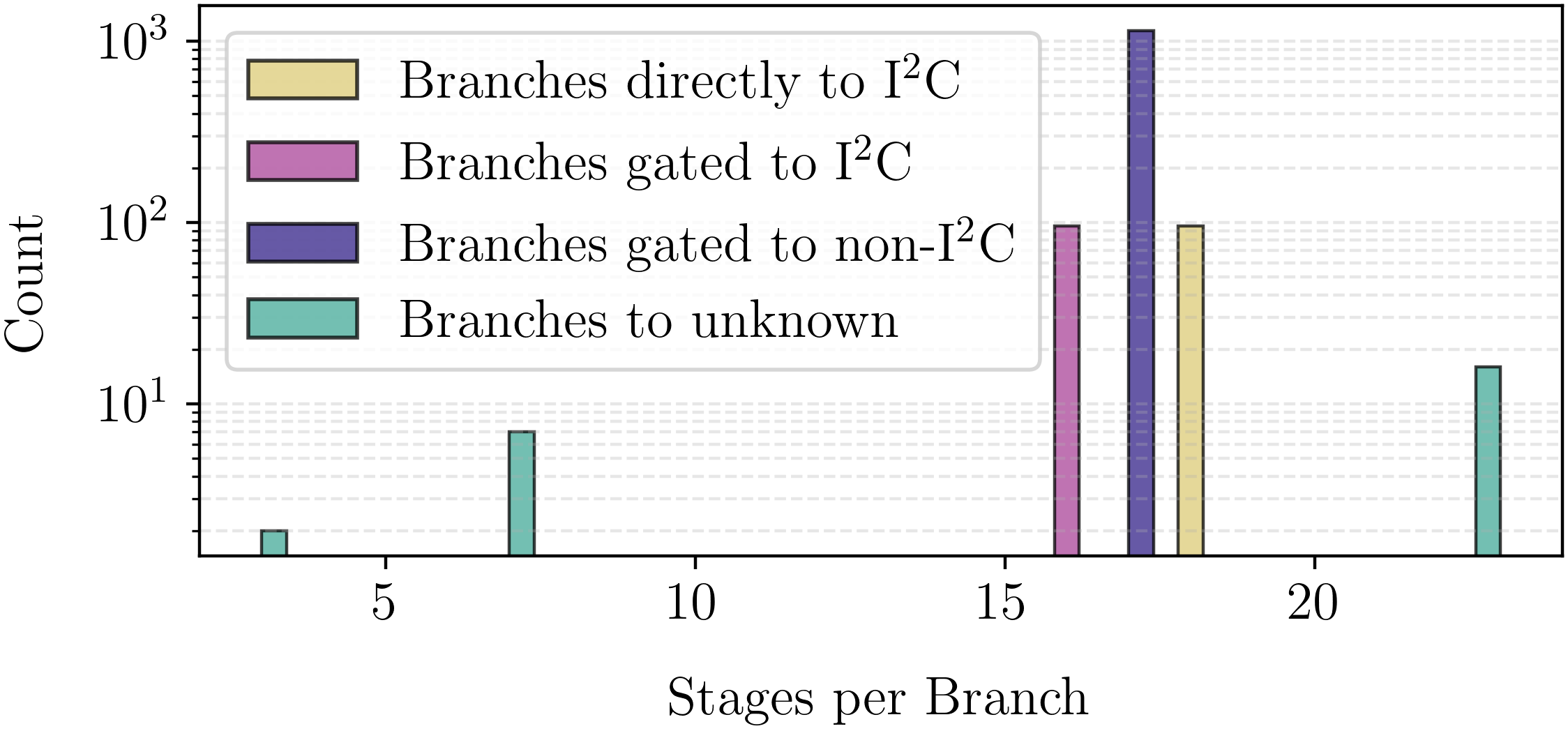}
\caption{
Number of stages per branch within the recovered \ac{CT}\@.
}
\label{fig:stages_per_branch}
\vspace{-8pt}
\end{figure}

On its way from the root to a sink, the clock traverses different numbers of gates.
When logic depth is equal across branches, active delays are nearly equivalent; unequal depths translate directly into clock skew.
Matching the logic depth across branches is thus the primary means of balancing their active delays~\cite{wu_effective_1992}.
The recovered \ac{CT} exhibits exactly this balance: the distribution of stages per branch is unimodal (see \autoref{fig:stages_per_branch}).
The majority of branches contain 17 stages, whereas only a few exhibit 16 or 18 stages.
Branches with 23, 7, or 3 stages are rare, isolated outliers unrelated to the special structures identified in \autoref{sec:control}.
That over a thousand sinks scattered across the die are reached through virtually identical logic depths is no accident: paths that would naturally be short carry additional inserted stages that match them to the rest.

The clock paths to the I\textsuperscript{2}C sinks are considerably shorter than those to the sinks of the H-tree backbone.
Left uncorrected, this disparity would cause the clock to arrive at the interface well before it arrives in the core---a skew between the two parts that must cooperate under normal operation.
Short paths cannot be shortened further, but they can be extended artificially.
This extension is realized exclusively with buffers, which compensate for the shorter clock paths to the I\textsuperscript{2}C sinks by introducing active delay through stage insertion, synchronizing the interface with the remainder of the digital core during normal operation.
The H-tree backbone, in contrast, needs to regenerate the signal along its long paths and relies exclusively on inverters---for driving, a typically economical choice, as it reduces area requirements and enhances robustness to duty-cycle fidelity~\cite{xanthopoulos_clocking_2009}.

\subsubsection{Groups Revealed by Routing}
\begin{figure}[htb]
\centering
\includegraphics[width=\linewidth]{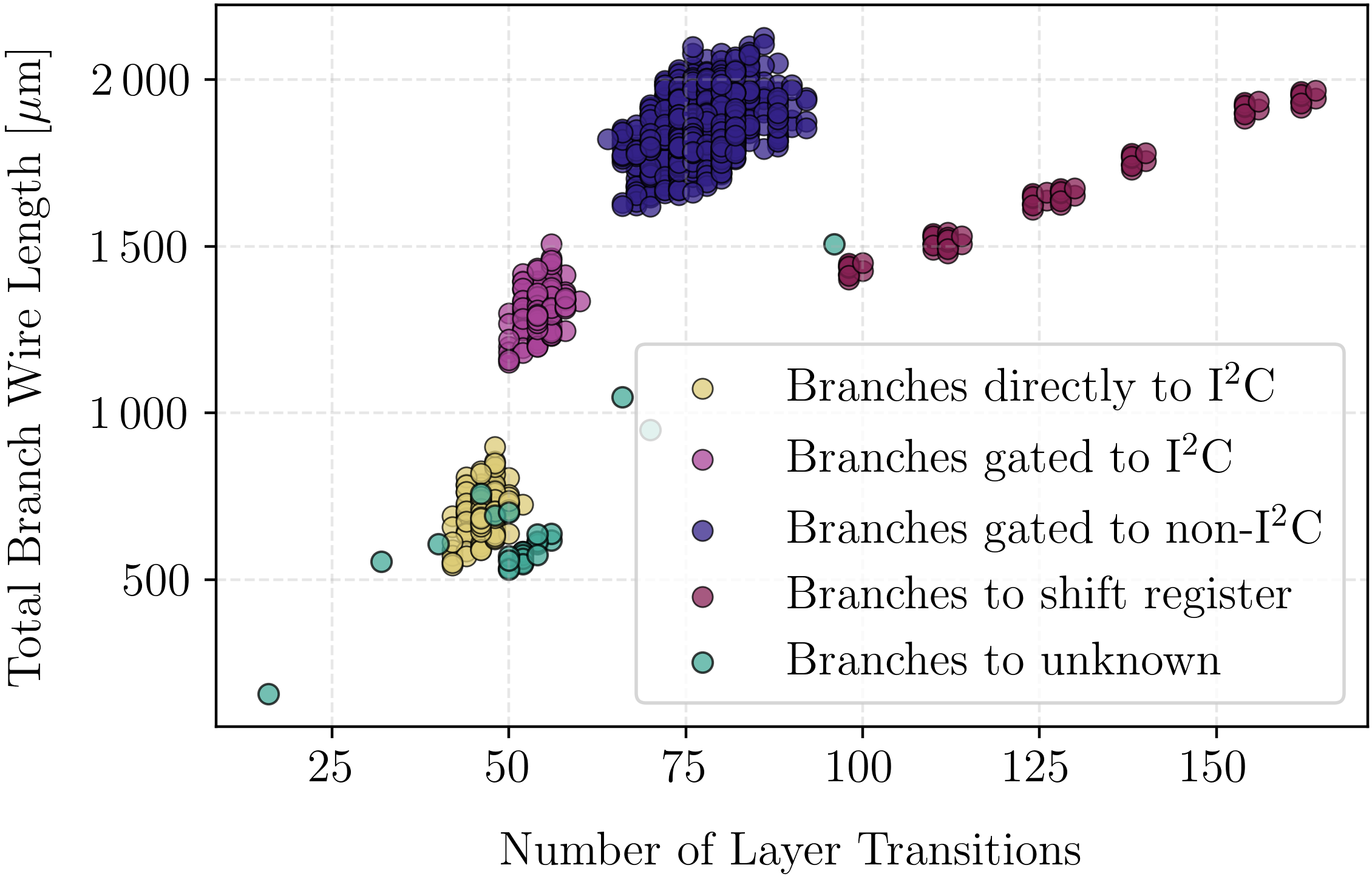}
\caption{
Total wire length per recovered \ac{CT} branch as a function of the number of metal-layer transitions\@.
}
\label{fig:segments_vs_length}
\end{figure}

Branches can be grouped not only by logic depth but by the delay implied by their metal routing. 
If the \ac{CT} is partitioned by function, branches within a group should share similar wire lengths and numbers of layer transitions---as observed in the data: per branch, total wire length correlates positively with the number of metal-layer transitions (i.\,e., via layers traversed), and the branches form distinct clusters (see \autoref{fig:segments_vs_length}).
Four dominant clusters appear between approximately 40 and 90 layer transitions, along with eight smaller clusters above 90 layer transitions.
Three of the four dominant clusters, as well as all eight smaller clusters, correspond to the structures identified in \autoref{sec:control}; each functional group of the \ac{CT} thus appears as its own cluster.

\subsubsection{Crosstalk Mitigation Strategies}
Our analysis of metal-track spacing in the recovered \ac{CT} routing shows no shielding guard tracks. 
While some regions have increased spacing, it is not applied consistently.

\subsection{Timing Analysis}
\label{sec:timing}
The structural analysis suggests deliberate design intent: balanced logic depths, inserted buffers, and routing partitioned by function.
Whether these measures achieve their goal is decided in nanoseconds, not in cell counts---by the time the clock actually arrives at each sink.
We therefore approximate the clock latency of every branch from the recovered geometry and its parasitics (cf.~\autoref{sec:timing_method}), obtaining the two quantities that no netlist contains: clock latency and clock skew.

\subsubsection{Clock Latencies}
\begin{figure}[htb]
\centering
\includegraphics[width=\linewidth]{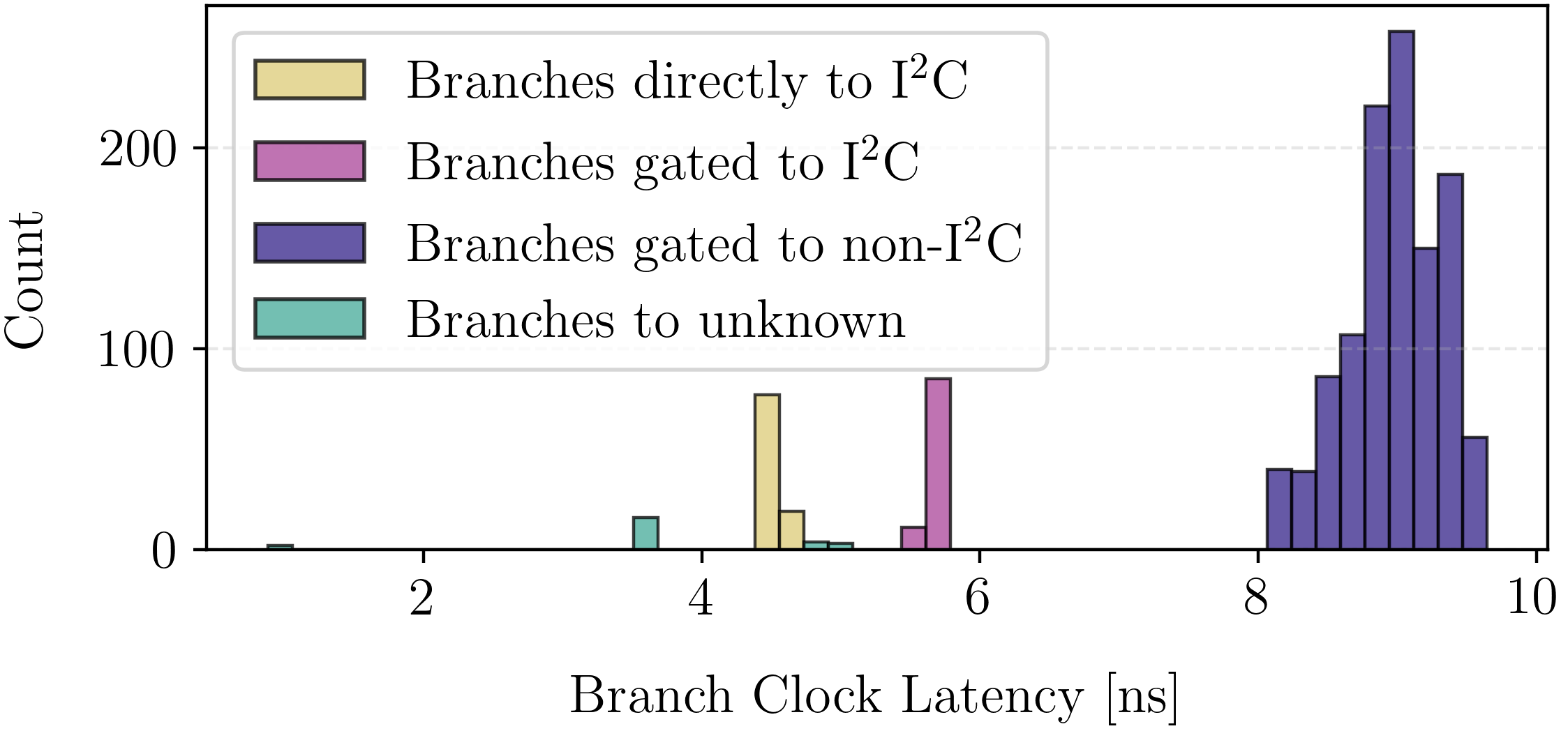}
\caption{
Clock latencies for relevant sinks of the recovered \ac{CT}\@.
}
\label{fig:clock_latency_distribution}
\end{figure}

We simulated clock latencies for all branches without clock-dividing structures, defining the clock at the first inverter after source selection. 
The routing clusters in \autoref{sec:silicon} indicate a functional partitioning of the recovered \ac{CT}, which the latency distribution supports by showing similar groupings (see \autoref{fig:clock_latency_distribution}). 
Direct I\textsuperscript{2}C branches show the lowest latencies ($\qty{4.39}{\nano\second}$ to $\qty{4.60}{\nano\second}$), followed by I\textsuperscript{2}C branches through the backbone enable ($\qty{5.59}{\nano\second}$ to $\qty{5.75}{\nano\second}$). 
All other backbone sinks have the highest latencies ($\qty{8.16}{\nano\second}$ to $\qty{9.64}{\nano\second}$).

\subsubsection{Clock Skews}
These latency ranges result in skews that may initially seem contradictory: while the global skew is substantial, each section of the \ac{CT} remains tightly balanced internally.
The timing behaviour of this \ac{IC} is characterized by four metrics: global skew ($\qty{8.77}{\nano\second}$), skew within the I\textsuperscript{2}C \ac{CT} ($\qty{0.21}{\nano\second}$), skew among H-tree-clocked sinks excluding those associated with the I\textsuperscript{2}C \ac{CT} ($\qty{1.49}{\nano\second}$), and skew across all H-tree-clocked sinks ($\qty{4.05}{\nano\second}$).
The substantial global skew primarily results from a few exceptionally short clock paths, an anticipated outcome of the partitioned \ac{CT}. 
The acceptability of a $\qty{8.77}{\nano\second}$ skew depends not on its absolute value but on its relationship to the clock period and critical paths.

This observation is consistent with a recurring pattern identified during the structural analysis: balancing was performed mostly within functional groups rather than globally, clock gating and switching were implemented using regular \acsp{MUX} and gates instead of glitch-free circuits, and crosstalk countermeasures were omitted.
This pattern suggests that timing was not under pressure and the designers economized accordingly.

%% file: figures/branches_overview.tex
\definecolor{clYellow}{HTML}{DDCC77} 
\definecolor{clGreen}{HTML}{AA4499}  
\definecolor{clPurple}{HTML}{332288} 

\tikzset{
  spdtA/.pic={
    \draw[thick] (0,-4.5mm) -- (-2mm,-0.6mm);
    \draw[fill=white] (-2.2mm,0)  circle (0.55mm) coordinate (-inA);
    \draw[fill=white] ( 2.2mm,0)  circle (0.55mm) coordinate (-inB);
    \draw[fill=white] (0,-4.5mm)  circle (0.55mm) coordinate (-out);},
  spdtB/.pic={
    \draw[thick] (0,-4.5mm) -- (2mm,-0.6mm);
    \draw[fill=white] (-2.2mm,0)  circle (0.55mm) coordinate (-inA);
    \draw[fill=white] ( 2.2mm,0)  circle (0.55mm) coordinate (-inB);
    \draw[fill=white] (0,-4.5mm)  circle (0.55mm) coordinate (-out);},
  spst/.pic={
    \draw[thick] (-2.5mm,0) -- (2.1mm,1.5mm);
    \draw[fill=white] (-2.5mm,0) circle (0.55mm) coordinate (-a);
    \draw[fill=white] ( 2.5mm,0) circle (0.55mm) coordinate (-b);}
}

\begin{figure}[htb]
\centering
\begin{tikzpicture}[
    font=\scriptsize,
    >={Latex[length=1.4mm]},
    sig/.style={->, semithick},
    wire/.style={semithick},
    ctl/.style={->, semithick, densely dashed},
    hl/.style={line width=2pt, opacity=0.75, line cap=round},
    ex/.style={line width=1.1pt, line cap=round},
    blk/.style={draw=black!35, rounded corners=6pt, align=center,
                minimum height=5.5mm, inner sep=2pt},
    src/.style={blk},
    lab/.style={font=\scriptsize, align=center}
]
\node[src] (sel) {select};
\node[src, right=3mm of sel] (osc) {on-chip clock};
\node[src, right=3mm of osc] (ext) {ext.\ clock (I/O)};

\path (osc.south) -- (ext.south) coordinate[midway] (m1);
\pic (rsw) at ($(m1)+(0,-6mm)$) {spdtA};
\node[lab, anchor=east] at ($(rsw-out)+(-6mm,2.5mm)$) {clock-source\\select};
\draw[sig] (osc.south) -- ++(0,-2mm) -| ([yshift=0.8mm]rsw-inA);
\draw[sig] (ext.south) -- ++(0,-2mm) -| ([yshift=0.8mm]rsw-inB);
\draw[ctl] (sel.south)  |- ($(rsw-out)+(-4mm,2.5mm)$);

\pic (esw) at ($(rsw-out)+(22mm,-5mm)$) {spst};
\draw[wire] (rsw-out) |- coordinate[pos=0.75] (tapU) ([xshift=-0.6mm]esw-a);
\filldraw (tapU) circle (0.6mm);
\draw[ctl, align=center] ($(esw-a)!0.5!(esw-b)+(0,6mm)$) node[above]{backbone\\enable} -- ++(0,-4mm);

\node[blk, below=20mm of rsw-out, xshift=12mm] (ht) {H-tree backbone};
\node[blk, right=5mm of ht] (xt) {X-tree};
\node[blk, minimum width=10mm,
      left=10mm of ht] (i2c) {I\textsuperscript{2}C};

\pic (isw) at ($(i2c.north)+(0,7mm)$) {spdtB};
\node[lab, anchor=west] at ($(isw-out)+(6mm,1mm)$)
      {I\textsuperscript{2}C clock\\select};
\draw[ctl, align=center] ($(isw-out)+(6mm,1mm)$) -- ++(-4mm,0);
\draw[sig] ([yshift=-0.8mm]isw-out) -- (i2c.north);

\coordinate (turn) at ($(ht.north)+(0,9mm)$);
\draw[wire] ([xshift=0.6mm]esw-b) -- ++(3mm,0) |- (turn);
\filldraw (turn) circle (0.6mm);

\draw[ex, clPurple, ->] (turn) -- (ht.north);
\draw[ex, clPurple, ->] (ht.east) -- (xt.west);
\draw[ex, clGreen]  (turn) -| ([yshift=0.8mm]isw-inB);   
\draw[ex, clYellow] (tapU) -- ++(0,-3.3mm)
                    -| ([yshift=0.8mm]isw-inA);           

\draw[hl, clYellow] (rsw-out) |- (tapU);
\draw[hl, clGreen]  ([xshift=1mm]rsw-out) |- ([yshift=-1.2mm]esw-a);
\draw[hl, clGreen]  ([yshift=-1.2mm]esw-b) -- ([xshift=2.4mm, yshift=-1.2mm]esw-b)
        |- ([yshift=1.2mm]turn);
\draw[hl, clPurple] ([xshift=-1mm]rsw-out) |- ([yshift=1.2mm]esw-a);
\draw[hl, clPurple] ([yshift=1.2mm]esw-b) -- ([xshift=4.8mm, yshift=1.2mm]esw-b)
        |- ([yshift=-1.2mm]turn);

\begin{scope}[shift={($(i2c.south west)+(-6mm,-4mm)$)}]
\foreach \x/\y/\col/\lab in {
    0/0/clYellow/{Branches directly to I\textsuperscript{2}C},
    38mm/0/clPurple/{Branches gated to non-I\textsuperscript{2}C},
    0/-3.6mm/clGreen/{Branches gated to I\textsuperscript{2}C}}{
  \draw[hl, \col] (\x,\y) -- ($(\x,\y)+(4mm,0)$);
  \node[anchor=west, font=\fontsize{7}{8}\selectfont] at ($(\x,\y)+(4.5mm,0)$) {\lab};}
\end{scope}
\end{tikzpicture}
\caption{Clock paths from inputs to sinks, with branches and control structures highlighted.}
\label{fig:architecture}
\end{figure}

%% file: sections/05_discussion.tex
\section{Discussion and Future Work}

Our case study demonstrates that layout-aware \acf{CTR} and \acf{CTA} provide insights not accessible from the \acf{GLN} alone.  
The evaluation targets an \ac{IC} in a mature $\qty{450}{\nano\meter}$ process with three metal layers.
Therefore, broader validation across designs and technology nodes is needed to establish generality. 
The \ac{CTR} currently targets single-driver nets and tree-like \acfp{CDN}; supporting meshes, clock-gating cells, and more complex clock dividers and multipliers would broaden the recoverable design space.

Timing analysis may be further refined by incorporating target-specific cell characterization, accurate clock-gate modeling, consideration of crosstalk and inductive effects, and empirical validation of resistance and capacitance estimates. 

Beyond competitive analysis, we envision the following applications for our methodology in hardware assurance.

\paragraph{Layout-Aware Netlist Analysis}
The broader use of layout information for \ac{HRE} remains largely unexplored. 
Physical features, including cell placement and routing characteristics, offer structural information absent from \ac{GLN}-only analysis and may enhance algorithms for recovering word-level structures, such as DANA~\cite{albartus_dana_2020}. 
More broadly, exploiting layout information for \acf{NA} could improve the recovery of higher-level design structures.

\paragraph{Trojan Detection}
Hardware Trojans may target the \ac{CDN} to manipulate clock frequency, introduce glitches, or launch fault attacks~\cite{karri_trustworthy_2010}. 
For instance, a Trojan may disable the clock or increase skew in specific functional components. 
Such modifications often manifest as anomalous clock-tree structures or clock gates. 
Consequently, \ac{CTR} could support detection of these Trojans, including electromagnetic-based Trojans such as the one described by Takou et al.~\cite{takou_optimal_2023}.

\paragraph{IP-Core Detection}
Modern \acp{SoC} frequently integrate \ac{IP} cores as semi-independent functional blocks~\cite{naik_integration_2021}. 
Because these blocks often contain dedicated \acp{CT} integrated into the central \ac{CDN} as subtrees, recovering \ac{CT} structures may offer an effective approach for localizing and identifying third-party \ac{IP}.

Together, these directions show that the \ac{CDN} is a rich and largely untapped source of design intent.
By recovering it from a fabricated \ac{IC}, our work opens the layout, not just the netlist, as a target for reverse engineering and assurance.

%% file: sections/acknowledgement.tex
\section*{Acknowledgment}
We thank Dr. Norman Fong for his analysis of the cross-sectional images, Zoltan Manyoki for his contributions to RC-delay modeling, and Aly Ahmed, Cynthia Mar, and Keith Doherty for their work on netlist recovery and partial circuit analysis, as well as the rest of the team at TechInsights.
This work was supported by the Deutsche Forschungsgemeinschaft (DFG, German Research Foundation) under Germany’s Excellence Strategy -- EXC 2092 CASA -- 390781972, and by the \href{https://rc-trust.ai}{Research Center Trustworthy Data Science and Security}, one of the Research Alliance Centers within the \href{https://uaruhr.de}{UA Ruhr}.